# Hybrid model for the analysis of human gait: a non-linear approach

João P. Galdino[1], P. H. Figueirêdo[2], Ramón E. R. González[3,*], Juan Lombana[4], Yésica Moreno[5], Sara M. Segura[6] and Carlos A. Collazo[7]

[1,2,3]Departamento de Física, Universidade Federal de Pernambuco, Recife, Brasil.
[4,5,6,7]Grupo de Investigación en Ingenieria Biomédica, Ciencias Básicas y Laboratorios, Universidad Manuela Beltrán, Bogotá, Colombia.

*Abstract*—In this work, a generalization of the study of the human gait was made from already existent models in the literature, like models of Keller and Kockshenev. In this hybrid model, a strategy of metabolic energy minimization is combined in a race process, with a non-linear description of the movement of the mass center's libration, trying to reproduce the behavior of the walk-run transition. The results of the experimental data, for different speed regimes, indicate that the perimeter of the trajectory of the mass center is a relevant quantity in the quantification of this dynamic. An experimental procedure was put into practice in collaboration with the research group in Biomedical Engineering, Basic Sciences and Laboratories of the Manuela Beltrán University in Bogotá, Colombia.

## I. INTRODUCTION

Modern biomechanics arose with the studies of Archivald V. Hill on the transformation of heat from the mechanical work of muscular contractions [1]. The correlation of thermodynamic parameters with the heat transformed by the muscles led him to win the Nobel Prize in Physiology and Medicine in 1922. With the discovery of Hill, the links between the macroscopic biological systems and the universal characteristics of Physics were tightened. The complexity of human locomotion comes from the fact that there are diverse and varied interactions between the body and the environment. A simplification of these interactions could be that the chemical potential energy originating from the muscles and the elastic potential energy of the tendons and type of the muscular elasticity end up transforming into work and heat [2]. The cyclic contractions in the active muscles give rise to reaction forces of the soil along the lower extremities. The force resulting from the gravitational action and the strength of resistance accelerates and decelerates the mass center of the body. Studies concerning the mechanical efficiency of locomotion [3] show that for different constant speeds, the rapidity in which energy is transformed, determined by oxygen consumption and by external work, is different for walking and running, the energy cost is lower in the first.

When we deal with bipedal movement in particular, nonlinear effects are often observed, but most of the approximate models for this movement ignore such effects. These models can be reliable for primary studies but end up being insufficient in relation to the complexity of the movement, losing the naturalness and clinical accuracy of the locomotion.

### A. Reaction force: the hypothesis of non-linearity

A reasonable approximation, when describing the process of gait, is to suppose a cyclic pattern of corporal movements that are repeated at each step [4]. The process of normal gait is one in which the erect body in movement is consistent with one leg and then on the other. This process has two phases, the support phase, which begins when the leg is in contact with the ground and lasts until this contact is lost. This phase represents 60% of the entire cycle of the march. The remaining 40% are the balance or balancing phase, since the leg loses contact with the ground until they come into contact again.

In the process of walking there are two very important basic aspects, the continuous reaction forces of the ground that sustain and provide the body with torque and the periodic movements of the feet from one position of support to the next, in the direction of movement. The reaction force of the soil depends intimately on the speed of travel [5]. The entire process is controlled by the neuromuscular system, which is why it is a complex process, which means that locomotion is a system in which changes in small components result in significant changes.

This work has received financial support from CAPES and CNPq (Brazilian Federal Grant Agencies) and from FACEPE (Pernambuco State Grant Agency)

[1]joao.pgaldinop,[2]phugof@gmail.com.
[3]ramayo_g@yahoo.com.br,[4]juan_lombana@hotmail.com.
[5]ing.moreno.r@hotmail.com,[6]sara.seguro@umb.edu.co,
[7]cacollazos@gmail.com
*Corresponding author.

### B. Metabolism and mechanical power for walking. The Keller model

People walk naturally in a way that energy consumption is optimized [4]. The metabolic rate, the variation of energy per time of physical activity, is generally measured indirectly by the amount of oxygen consumed during bodily activity [6]. In order to minimize energy dissipation, the neuromuscular system "selects" the speed of the individual. Deviations from this normal gait pattern result in increased energy consumption and limit locomotion [7].

In 1974, J. B. Keller [8] proposed a model based only on variational calculus and elementary dynamics to study the career of extraordinary human performance. In this model, it is assumed that the reaction force of the soil affects the amount E (t) of power per unit mass from muscle stores that store N_2 of the food and the consumption of O_2 of the individual. The reactions that occur in the body of the individual use these chemical reserves to provide power for locomotion. For prolonged periods of activity, the individual's biology supplies a σ rate of energy supply obtained from respiration and circulation. Keller determined the theoretical relationship between the shortest time T, based on physiological parameters, in which a distance D can be traversed (1).

$$D = \frac{F}{\gamma^2}(\gamma T - 1 + e^{-\gamma T}) \tag{1}$$

In this calculation, it was considered that the propulsive force of the sprinter cannot exceed a certain maximum value (2).

$$f(t) \leq F \tag{2}$$

The constant $\gamma$ is a physiological constant and has units of time inverse. So that there is an optimized solution, according to the results of Keller's work, for the sprinter in starting a finite number of velocity curves described according to (3) is obtained.

$$v(t) = \left(\frac{\sigma}{\gamma} + Ce^{-2t\gamma}\right)^{\frac{1}{2}}, \tag{3}$$

where the constant C is arbitrary and is determined by the initial velocity.

The important cases in Keller's work are: an acceleration curve during the interval in which the runner exerts the maximum momentum; the constant velocity curve from the moment the runner reaches its maximum speed and the deceleration curve that begins when $E(t) \equiv 0$.

These three curves are combined ensuring continuity throughout the run and that the area under the speed curve must be maximum. A summary of the above can be seen in (4) and in figures 1 and 2.

$$v(t) = \begin{cases} \frac{F}{\gamma}(1 - e^{-\gamma t_1}), & 0 \leq t \leq t_1 \\ V, & t_1 \leq t \leq t_2 \\ \left(\frac{\sigma}{\gamma} + \left(V^2 - \frac{\sigma}{\gamma}\right)e^{-2\gamma(t-t_2)}\right)^{\frac{1}{2}}, & t_2 \leq t \leq T \end{cases} \tag{4}$$

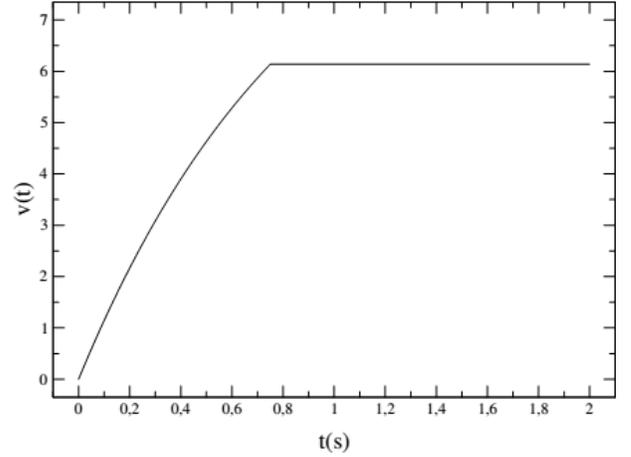

Fig. 1. Speed behavior of $D < D_c$, where $D_c$ is the critical distance. There is an impulse given by the variation of the velocity until the first transition, where, after the transition, the velocity remains constant and equal to $V$.

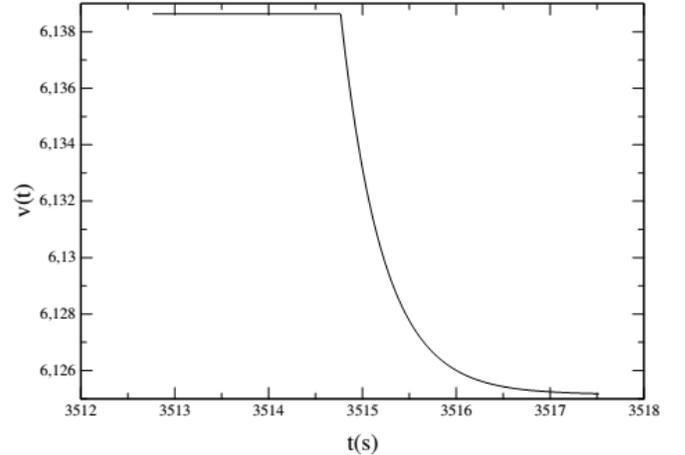

Fig. 2. Behavior of the speed when the energy power is carried to the limit $E(t) = 0$. In the transition, the curve presents a decline in speed referring to the athlete's energy limit.

When analyzing competitions between professional athletes and world record data. Keller estimated parameters for a run with optimal or next optimization strategy. These values, corresponding to physiological variables are the following (5).

$$F = 12.2 \frac{m}{s^2}, \frac{1}{\gamma} = 0.892 \, s, \sigma = 41.56 \frac{m^2}{s^3}, T_c = 27.7 \, s, E_0 = 2409 \frac{m^2}{s^3} \tag{5}$$

The parameters $T_C$ and $E_0$ are the critical time (the minimum for the distance D) and the initial power per unit mass.

### C. Kokshenev model for constant speed walks.

For the study of the oscillations that the center of mass experience in a locomotion at constant speed it is important to observe the movement of the center of mass in a sagittal plane. Kokshenev, in his 2004 paper [9], used this plane as a reference to study the movement of the center of mass during human walking at constant speed. In this model, an inertial reference system is considered moving as a virtual mass center defined by the displacement vector $\overrightarrow{R_0}(t)$. The conditions established in the model for displacement are: $x_0(t) = Vt$, where $V$ is a constant and $y_0(t) = H$ is the average height of the center of mass in relation to the ground, where the origin of the inertial coordinate system is. In this way, the displacement vector relative to the movement of the center of mass of the human body with the virtual center of mass is $\overrightarrow{\Delta r}(t) = \vec{R}(t) - \overrightarrow{R_0}(t)$, where $\vec{R}(t)$ is the displacement vector of the center of mass of the human body in relation to the same referential of $\overrightarrow{R_0}(t)$.

In the model, a driving force $\overrightarrow{\Delta F}(t)$ is defined, which is the neuromuscular capacity to exercise work, which is derived from observations of small oscillations close to the weight support of the body. In (6), we can see this force related to the ground reaction force $\vec{F}(t)$ and the force of gravity.

$$\vec{F}(t) = -m\vec{g} + \overrightarrow{\Delta F}(t) \qquad (6)$$

The driving force, as well as the velocity and displacement of the center of mass must respect the condition of cyclosymetry that preserves the amount of movement of the system.

Applying the Lagrangean formalism, and considering harmonic and non-harmonic solutions for the movement of the center of mass, Kokshenev found the force that characterizes the forced oscillatory movement of the center of mass around the point $\overrightarrow{\Delta r} = \overrightarrow{\Delta r_0}$ (7).

$$\vec{F}(t) = \overrightarrow{\Delta F_0} + \overrightarrow{\Delta F_1}(t) \qquad (7)$$

The first term of (6) is the force that describes the free movement of the center of masses as a superposition of linear oscillations.

With the increase of the speed, the anharmonic effects become important, being necessary the introduction of a potential resulting from the expansion in a Taylor series of the elastic potential of the Hamiltonian. The second then results from the gradient of this expanded potential up to the order of the anharmonic effects. Result of all this, the components of the force given by (6) are presented in (8) and (9).

$$\overrightarrow{F_x}(t) = -m\omega_0{}^2(\Delta l_0 \sin(\omega_0 t) - \Delta l_1 \sin(2\omega_0 t)) \qquad (8)$$

$$\overrightarrow{F_y}(t) = m\vec{g} + m\omega_0{}^2(\Delta h_0 \cos(\omega_0 t) - \Delta h_1 \cos(2\omega_0 t)) \qquad (9)$$

The coefficients $\Delta l_0(v), \Delta h_0(v), \Delta l_1(v)$ and $\Delta h_1(v)$ are harmonic and anharmonic amplitudes respectively, whose values correspond to experimental data and $\omega_0(v)$ corresponds to the frequency of a cycle on the step cycle. Finally, the introduction of a locomotive resistive force $\overrightarrow{\Delta F_{res}}(t) = -\gamma \overrightarrow{\Delta r_1}(t)$ where $\gamma(v)$ represents the coefficient of friction, results in the following functions for the respective positions in a steady state with $\omega_0 t \gg 1$ and in a low resistance approach (10) and (11).

$$x(t) = Vt + \Delta x_0(t) + \frac{\Delta l_1}{3} \frac{\sin(2\omega_0 t + \vartheta)}{\sqrt{1 + \tan(\vartheta)^2}} \qquad (10)$$

$$y(t) = H + \Delta y_0(t) + \frac{\Delta h_1}{3} \frac{\cos(2\omega_0 t + \vartheta)}{\sqrt{1 + \tan(\vartheta)^2}} \qquad (11)$$

These equations are the solution of the following equation of motion (12).

$$m\overrightarrow{\ddot{\Delta r_1}}(t) + \gamma\overrightarrow{\dot{\Delta r_1}}(t) + k_0\overrightarrow{\Delta r_1}(t) = \overrightarrow{\Delta F_1}(t) \qquad (12)$$

The force $\overrightarrow{\Delta F_1}(t)$ is given by (7) and (8) previous.

The results of Kokshenev show a closed orbit given by a hypocycloid $(\Delta r_1 < \Delta r_0 \ll H)$, around a fixed point and clockwise. It is assumed, in the work of Kokshenev, that this orbit is described, for walking, as a characteristic ellipse, with amplitudes $\Delta l = \Delta l_0 + \frac{\Delta l_1}{3} \frac{1}{\sqrt{1 + \tan(\vartheta)^2}}$ and $\Delta h = \Delta h_0 + \frac{\Delta h_1}{3} \frac{1}{\sqrt{1 + \tan(\vartheta)^2}}$, horizontal and vertical, respectively. Given the conditions of the model, the center of mass moves with constant speed $V$ at a certain height $H$ and rotates along a hypocycloid circumscribed by a "flattened" or "shrunken" ellipse of eccentricity $e_+(e_-)$ given by the following expression (13).

$$e_\pm = \sqrt{1 - \left(\frac{\Delta l_0}{\Delta h_0}\right)^{\pm 2}} \qquad (13)$$

## II. EXPERIMENTAL ANALYSIS OF WALK FOR DIFFERENT SPEEDS

### A. Experimental environment and appliances

Experimental data were collected in a space of approximately $16 \, m^2$, where there is a track formed by four force platforms and six motion detection cameras around the platforms, see in figure 3.

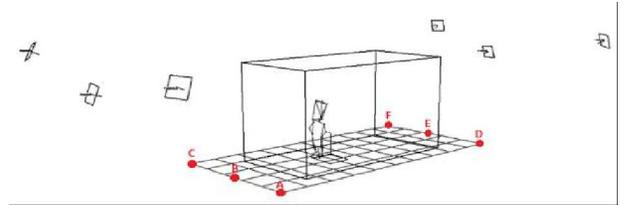

Fig. 3. Scheme that shows the space where the experiments were conducted. The figure portrays the volume occupied by the track and platforms. The motion detection cameras were distributed in the positions of points A, B, C, D, E and F, selected to facilitate the orientation of the volunteers.

The cameras used are part of the data acquisition system for BTS GAITLAB motion analysis. These optoelectronic cameras measure the displacement, with an accuracy of $\pm 10^{-7}$ m, of body segments in a time interval of $\pm 10^{-2} s$ [10].

*B. Positioning protocol for markers and orientation of body segments. Location of the center of masses.*

The Davis protocol was used for the placement of markers distributed in different segments and corporal regions of the volunteer (figure 4). For each range of speed (around 1.03 m/s, 1.81 m/s and 3.2 m/s), we counted on the records of five volunteers of the feminine sex and six male volunteers.

In literature, we can find several references in relation to the position of the center of mass of the human body. For Miralles [11] the center of mass lies behind the lumbar vertebra L5. Yet for Dufour and Pillu [12] it is located before the sacral vertebra $S_2$. We suggest in this work that the center of mass would be placed just between these two vertebrae. The Smart TRACKER and Smart Analyzer was used, a program with which we is able see the track and to capture the position of each of the markers placed on the volunteer. A simple model was created that virtually simulates the markers and their connection (figure 4).

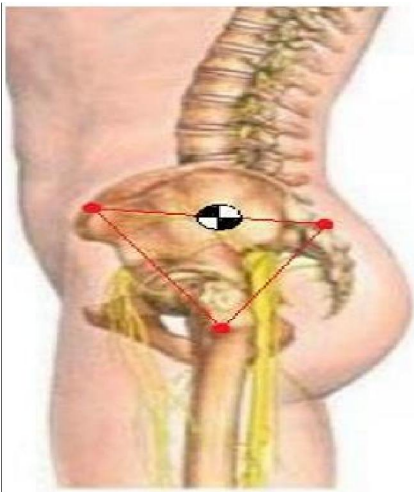

Fig. 4. Position of the center of masses. Midpoint between the referenced locations (by the authors).

With the Smart TRACKER and Smart ANALYZER, the displacements of the real markers coupled with the volunteer was interpolated, guaranteeing the continuity of the information throughout the capture of the data. With the defined function, a virtual point situated at the midpoint between the iliac crests and sacrum was created (4, 5 and 6 markers in figure 5). On this point all the clinical analysis corresponding to the march was done.

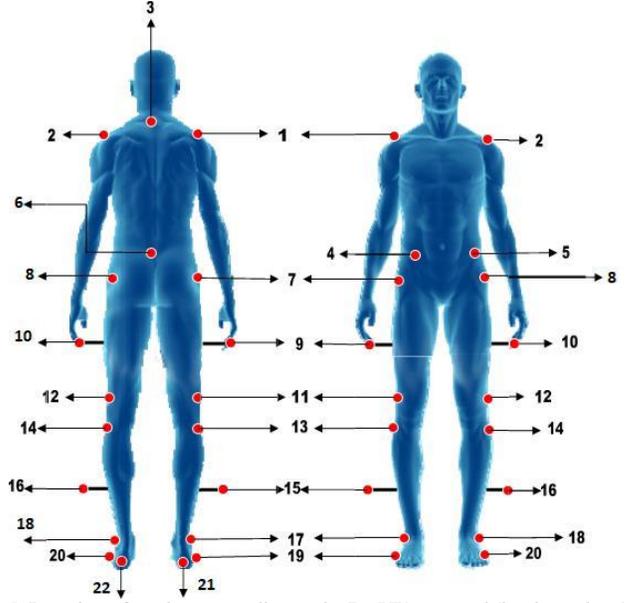

Fif. 5. Location of markers according to the DAVIS protocol (by the authors).

## III.  HYBRID MODEL FOR THE WALKED-RUN TRANSITION

*A. Details of the model*

In order to find results for different walking speed using the Kokshenev model for non-linear running, we used the experimental data conceived in the experiments described above for comparison with important aspects of the model. In our model we use (12), deduced in the Kokshenev model, which, from a simpler mechanical analog, represents oscillations in the two-dimensional plane under the action of a viscous resistant force and the reaction force of the soil acting on favor of the movement, turning the system a two-dimensional pendulum damped and forced.

The diversity of the velocities was issued by (4). The physiological parameters used in optimization by Keller were adopted (5). We noted that, by varying these parameters and implementing them in the equation for speed, there was a dependence between the terms related to the transitions between the different speed regimes. At time $t_1$, which separates the velocity regime with exponential growth of the constant velocity regime, the following behavior is obtained (14).

$$t_1 = -\frac{1}{\gamma} ln\left(1 - \frac{v\gamma}{F}\right) \tag{14}$$

From this, a maximum value is induced for the ratio F/γ from which we can find $t_1$. In this way, the relationship between the parameters F and γ is established as the maximum limit for the value of the speed reached (maximum speed). From the previous equation (14) we can see that the maximum speed that the system can reach is equal to the limit in which time tends to infinity in the following relation (15).

$$v(t) = \frac{F}{\gamma}(1 - e^{-\gamma t}) \tag{15}$$

Making the maximum speed can be expressed according to (16).

$$v_{max} = \frac{F}{\gamma} \qquad (16)$$

The minimum speed, on the other hand, refers to the case in which the system works for long periods of time. At the time $t_2$ when the physiological wear begins, the velocity transition occurs and this time we can obtain it from (4), as follows (17) [woodside].

$$t_2 = \frac{\left\{ E_0 + \frac{FV}{\gamma} + \left[ \left( \frac{F}{\gamma} \right)^2 - V^2 \right] \ln\left[ 1 - \left( \frac{V\gamma}{F} \right) \right] \right\}}{(V^2 \gamma - \sigma)} \qquad (17)$$

In order to maintain the coherence of the function from t_2 to T, in case that v is less than the root of the term $\sigma/\gamma$, then we will have a divergence on the function that characterizes this regime transition. We can then interpret the following as the minimum speed (18).

$$v_{min} = \sqrt{\frac{\sigma}{\gamma}} \qquad (18)$$

For the values of the physiological parameters cited, the maximum time T that an individual can reach, at the end of the slope curve due to physiological wear is (19).

$$T = t_2 - \frac{1}{2} \ln\left\{ \frac{1}{4} \left[ 1 - \left( \frac{\sigma}{V^2 \gamma} \right) \right] \right\} \qquad (19)$$

In this equation, we can see that for values of $\sigma \ll V^2 \gamma$ there will be a value of $T$ while for $\sigma \gg V^2 \gamma$ the equation diverges, where $T$ is in that case inaccessible. The connection between the parameters $\sigma$ and $T$, as well as the minimum speed allows us to find different values of $\sigma$ respecting the maximum and minimum values of the speed.

In (12), the terms $k_0$ and $\gamma$ are functions of velocity. The form as $\omega_0(v) = \frac{k_0(v)}{m}$ varies was obtained by Kokshenev [9] using the results of experimental data reported in [3]. It was defined that $\omega_0(v)$ is a linear function of velocity, as follows (20).

$$\omega_0(v) = 4.94 + 4.02v \qquad (20)$$

With this result, the coefficient of friction per unit of mass $\omega_1(v) = \frac{\gamma(v)}{m}$ was found as follows (21).

$$\omega_1(v) = 6.37 - 6.15v + 2.38v^2 \qquad (21)$$

Introducing Keller's optimal speed and substituting the dependence with speed for the dependence with time, for small oscillations we can affirm that (22).

$$\omega_0(t) \approx 4.94 + 4.02\left( 1 - e^{-v(t)} \right) \qquad (22)$$

In the numerical solution (12), using (20) and (21) e with $v(t)$ being Keller's optimal velocity, we note that it grows rapidly for the constant velocity value and (12) is quickly damped by the growth of $\omega_1(v)$. The model presented here never comes to contemplate the physiological wear due to the quadratic growth of $\omega_1(v)$ and its mathematical complexity.

The physiological parameter chosen to determine the minimum velocity was $\sigma$. For each value of $\sigma$, the last oscillation of the $x$ and $y$ positions was recorded and parametric curves were constructed. These curves vary between the maximum time $t_{max}$ of the occurrence of the movement and the difference between that time and an $R$ term dependent on the angular velocity $\omega_0$ in $t_{max}$ as follows (23).

$$R = \frac{2\pi}{\omega_0(t_{max})} \qquad (23)$$

The registration time of a parametric curve is (24):

$$t_{max} - R < t < t_{max} \qquad (24)$$

## IV. RESULTS

With the equations already defined, the behaviors of the components $x(t)$ e and $y(t)$ for the trajectory of the center of masses were determined. Three velocity values were chosen: v = 1.03 m/s, 1.81 m/s and 3.20 m/s, referring to those obtained in collaboration with the Research Group in Biomedical Engineering, Basic Sciences and Laboratories of the Manuela Beltrán University, in Bogotá, Colombia. The final period of oscillation and comparison with a period taken from the real data was plotted from the model. We chose the data closest to an average of a characteristic behavior of force, for the experimental data. The characteristic behavior of the normalized force of the actual data is related to the average behavior of the data obtained with all the volunteers and platforms. In this way, we selected the data of the individual MF7P1, female, on the $P_1$ platform, one of the four used to capture the FRS, see figures 6 and 7.

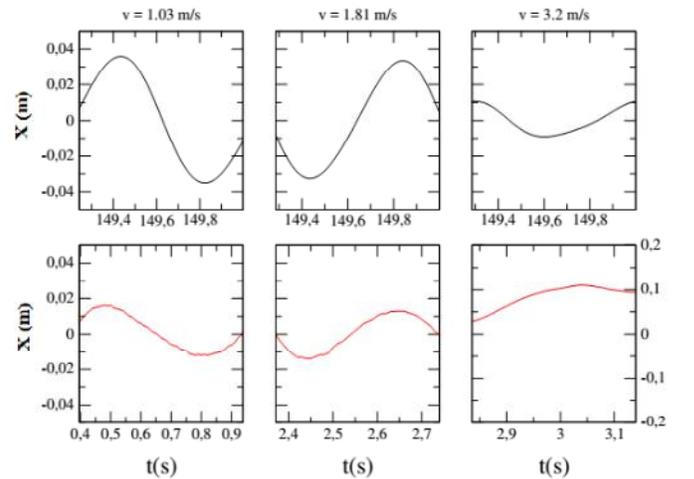

Fig. 6. Behavior of the position $x(t)$ of the center of masses for three different speeds in different periods. The black lines represent the results derived from the model and the red lines, the results from the real data. The first column on the left shows the graphs for the velocity v = 1.03 m/s. These graphs are in phase and the amplitude obtained with the model are twice as large. The middle column shows the results for the velocity v = 1.81 m/s. The curves are also in phase and the ratio between the amplitude is the same as for the lower speed. The third column, for v = 3.20 m/s, shows curves in phase opposition, in this case, the amplitude of the curve referring to the actual data is four times greater.

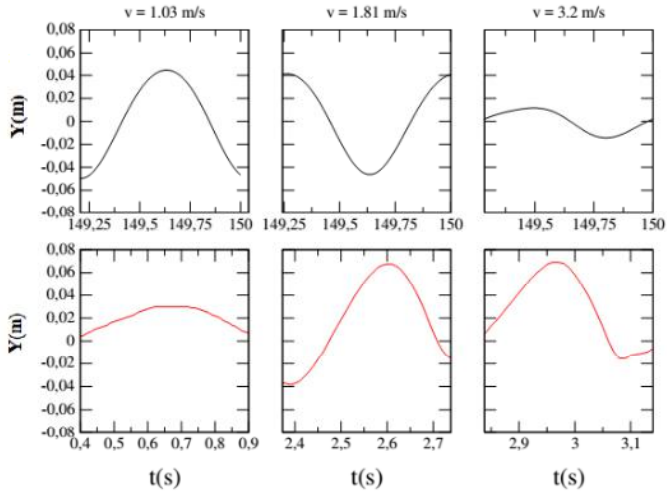

Fig. 7. Behavior of the position $y(t)$ of the center of masses for three different speeds in different periods. The black lines represent the results derived from the model and the red lines, the results from the real data. The first column on the left shows the graphs for the velocity v = 1.03 m/s, where a similar maximum point is seen in both graphs. The central column shows the results for v = 1.81 m/s and is seen that the curves have inverted phases. The column on the right represents the velocity v = 3.20 m/s, where it can be seen that the two curves oscillate in phase, although the amplitude of the curve referring to the actual data is three times greater.

We see in Figure 6, for the first two speed, that the oscillations have the same phase, both in time and in amplitude. Yet for the speed of 3.2 m/s, we see a phase shift between the curves of approximately $\pi / 2$ and the values of the amplitude referring to the real data are in a proportion four times greater than the amplitude of the model.

In Figure 7, the same time interval was plotted. It is observed, to 1.03 m/s a point of maximum similar in both graphs, of the model and of the real data. For the intermediate speed, 1.81 m/s we see extremes of inverted phases and for 3.20 m/s, although the amplitude obtained from the model is three times lower than that obtained from the actual data, the phases of the oscillations are quite similar.

From the model, parametric curves for x (t) and y (t) were obtained by varying the value of the physiological constant σ, reported by Keller (5) from 0.2 m²/s³ to 12 m²/s³. It was observed that there is a perimeter for the trajectory, associated with each translation speed and that the velocity value that maximizes the perimeter is v = 1.38 m/s.

Figure 8 illustrates the behavior of the perimeter of the trajectories of the center of masses for different translation speed. At speeds up to v = 1.38 m/s, the perimeter of the trajectories increases and higher speed it decreases.

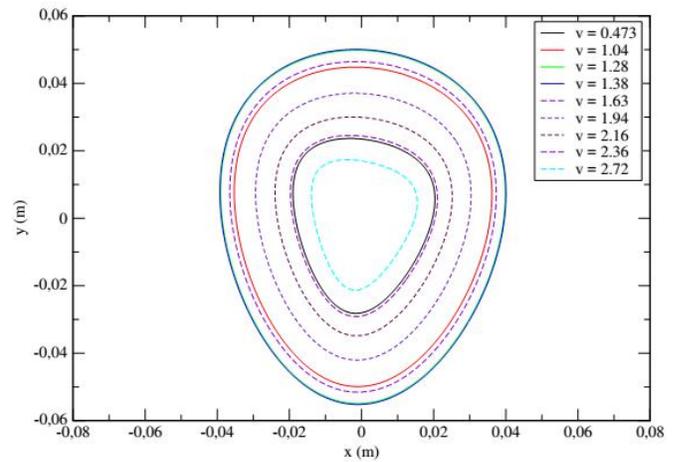

Fig. 8. Parametric curves of the trajectories of the center of masses for different translation speed. The solid lines represent the lowest speeds, for which a gradual increase in the perimeter is seen as a function of the speed up to v = 1.38 m/s. For higher speeds, represented by dashed lines, we see a decrease in the perimeter of the trajectories as the speed of translation increases, showing that the amplitudes in the curves become increasingly smaller with the increase in speed.

For each of the curves in Figure 8, the perimeter was calculated, corresponding to the trajectory of the center of mass for each velocity. A graph of the perimeter as a function of speed was constructed where the maximum point was easily identified, corresponding to the velocity v = 1.38 m/s. The perimeters corresponding to experimental data found in the literature were also calculated [13] and trajectories of the center of mass obtained from the experimental data, in collaboration with the Research Group on Biomedical Engineering, Basic Sciences and Laboratories of Manuela Beltrán University, in Bogotá, Colombia. These last data are referring to a certain running regime, with a definite velocity of the center of mass of the volunteer. With this data obtained an average of the trajectories of the center of mass of each volunteer and with such means a closed parametric curve was generated, for which its characteristic parameter was calculated. The perimeter of the path of just one individual was likewise calculated, which approximated the result obtained by the average in each speed regime.

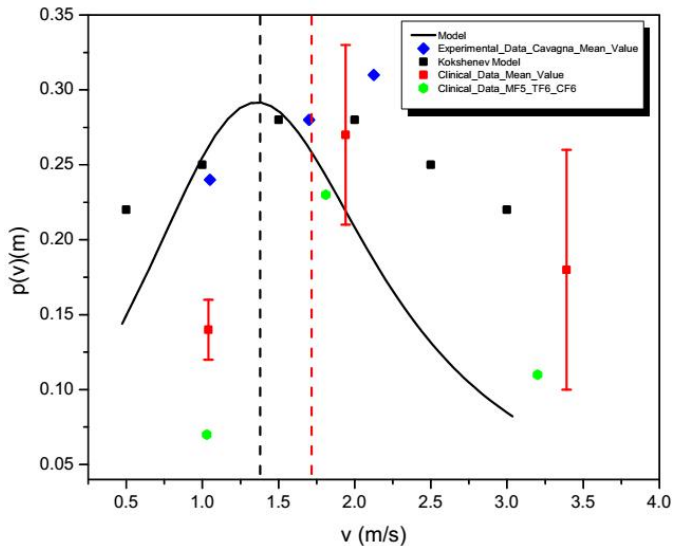

Fig. 9. Perimeter according to the translation speed. The line represents the points obtained from the model. The blue points were available from the literature [ref 22]. The red dots are referring to the average of the actual data, obtained from four volunteers. The green points are the perimeters of a volunteer that was felt better to the curve of the model. Finally, the black points are the perimeters corresponding to the eccentricities of the literature reported by Kokshenev [9].

In the graph of figure 9, we can see all the information regarding perimeters as a function of speed. From experimental results reported by Kokshenev, it was possible to find the perimeter corresponding to each plotted eccentricity, as a function of the speed and they were represented using black squares. With red squares the average values of the perimeters corresponding to the experimental data were represented in each speed regime. The blue points are referring to real data from the literature [14] and the green points are the perimeters of the trajectories of a single individual, MPF5P1, which is adapted to the values obtained for the average of the parametric trajectories. The dashed red line marks the speed considered by Kokshenev as the transition point between the march and the race, this speed is v = 1.73 m/s. The dashed black line represents the maximum perimeter velocity v = 1.38 m/s. This speed represents, for our model, what would be the transition between the running and running regimes. It is reasonable to think that close to this point of maximum of the gait is located, said in some way, normal for a healthy adult [15]. This walking speed is the most stable for the mass center of healthy adults.

The proposed model presents peculiar behavior in relation to the perimeter of the trajectory of the center of masses as a function of the translation speed. Trying to solve the model for variable speeds, in the way proposed by Keller, has ineffective implications, since the translation speed has a very short duration in the regime in which the speed grows exponentially, observed in equations (4) and lasts a long time in the stationary speed regime, before reaching physiological wear. In fact, for practical purpose, physiological wear is unattainable for the simulation time that is generated.

Kokshenev work with the hypothesis that the eccentricity varies depending on the speed. In this work we chose to study the perimeter depending on the speed as an approach to obtain the results. This approach is plausible due to the relationship between the eccentricity and the amplitude of the curves. The visibility between the perimeter and the speed was another point for the use of this approach. A result of this relationship is the maximum point found, for v = 1.38 m/s. This value coincides with one of the most accepted values in the literature for the "normal" speed of a healthy adult [ref 31 and 32].

For the topic addressed in this paper there are various methods and models in the literature [ref 10, 26, 33 and 34], on the other hand, the model studied here, despite being a simplification of effects of other natures, is acceptable because, using only non-linear mechanics, the results obtained result in a good approximation of reality.

## V. CONCLUSIONS

The objective of this work was to approach non-linear effects in biomechanics. Using models already known from the literature, such as Kokshenev and Keller, an association of these models was achieved in order to obtain results from a model with more general characteristics. An experimental procedure was adopted in collaboration with the research group in Biomedical Engineering, Basic Sciences and Laboratories of the Manuela Beltrán University in Bogotá, Colombia. These experiments generated data that, finally, were compared with the created model.

We saw that, for low speeds, the proposed model works quite well, from v = 1.0 m / s to approximately v = 1.38 m / s, for which the perimeter of the center of mass calculated from the model coincides or results fairly close to the experimental perimeter. As the speed increases, deviations are observed more and more accentuated. It is also observed that the trajectory is more accentuated for the critical speed v = 1.38 m / s and decreases both, with the increase and with the decrease in speed.

It is noticeable that the presented model manages to approximate the experimental results for low speed of the march, in spite of the model using physiological parameters that optimize the gait. Deviations at high speed are hypothetically associated with noise from the central pattern generator (PCG), a biological neural network responsible for locomotion that produces a rhythmic pattern in the absence of sensory responses or descendants that carry specific temporal information [ref 35].